
\input harvmac
\noblackbox


\def\bfone{\relax{\rm 1\kern-.35em 1}}
\def\inbar{\vrule height1.5ex width.4pt depth0pt}

\def\IC{\relax\,\hbox{$\inbar\kern-.3em{\rm C}$}}
\def\ID{\relax{\rm I\kern-.18em D}}
\def\IF{\relax{\rm I\kern-.18em F}}
\def\IH{\relax{\rm I\kern-.18em H}}
\def\II{\relax{\rm I\kern-.17em I}}
\def\IN{\relax{\rm I\kern-.18em N}}
\def\IP{\relax{\rm I\kern-.18em P}}
\def\IQ{\relax\,\hbox{$\inbar\kern-.3em{\rm Q}$}}
\def\us#1{\underline{#1}}
\def\IR{\relax{\rm I\kern-.18em R}}
\font\cmss=cmss10 \font\cmsss=cmss10 at 7pt
\def\ZZ{\relax\ifmmode\mathchoice
{\hbox{\cmss Z\kern-.4em Z}}{\hbox{\cmss Z\kern-.4em Z}}
{\lower.9pt\hbox{\cmsss Z\kern-.4em Z}}
{\lower1.2pt\hbox{\cmsss Z\kern-.4em Z}}\else{\cmss Z\kern-.4em
Z}\fi}

\def\cN{{\cal N}} 
\def\cP{{\cal P}} 
\def\cR{{\cal R}} \def\cS{{\cal S}}
 \def\cV{{\cal V}}
\def\nup#1({Nucl.\ Phys.\ $\us {B#1}$\ (}
\def\plt#1({Phys.\ Lett.\ $\us  {B#1}$\ (}
\def\cmp#1({Comm.\ Math.\ Phys.\ $\us  {#1}$\ (}
\def\prp#1({Phys.\ Rep.\ $\us  {#1}$\ (}
\def\prl#1({Phys.\ Rev.\ Lett.\ $\us  {#1}$\ (}
\def\prv#1({Phys.\ Rev.\ $\us  {#1}$\ (}
\def\mpl#1({Mod.\ Phys.\ Let.\ $\us  {A#1}$\ (}
\def\ijmp#1({Int.\ J.\ Mod.\ Phys.\ $\us{A#1}$\ (}
\def\jag#1({Jour.\ Alg.\ Geom.\ $\us {#1}$\ (}
\def\tit#1|{{\it #1},\ }

\def\Coe#1.#2.{{#1\over #2}}

\def\coe#1.#2.{\relax{\textstyle {#1 \over #2}}\displaystyle}

\def\sqr#1#2{{\vcenter{\vbox{\hrule height.#2pt
         \hbox{\vrule width.#2pt height#1pt \kern#1pt
            \vrule width.#2pt}
         \hrule height.#2pt}}}}
\def\square{\mathop{\mathchoice\sqr56\sqr56\sqr{3.75}4\sqr34}\nolimits}

%
\def\nihil#1{{\it #1}}
\def\eprt#1{{\tt #1}}
\lref\NWcpts{N.P.\ Warner, \nihil{Some New Extrema of the Scalar
Potential of Gauged $N=8$ Supergravity,}
Phys.~Lett.~{\bf 128B} (1983) 169.}
\lref\GRW{M.\ G\"unaydin, L.J.\ Romans and N.P.\ Warner,
\nihil{Gauged $N=8$ Supergravity in Five Dimensions,}
Phys.~Lett.~{\bf 154B} (1985) 268; \nihil{Compact and Non-Compact
Gauged Supergravity Theories in Five Dimensions,}
Nucl.~Phys.~{\bf B272} (1986) 598.}
\lref\PPvN{M.~Pernici, K.~Pilch and P. van Nieuwenhuizen,
\nihil{Gauged $N=8$ $D = 5$ Supergravity,}
Nucl.~Phys.~{\bf B259} (1985) 460.}
\lref\CNPNW{C.N.\ Pope and N.P.\ Warner, \nihil{An $SU(4)$
Invariant  Compactification of $d = 11$ Supergravity on a Stretched
Seven Sphere,} Phys.~Lett.~{\bf 150B} (1985) 352;
\nihil{Two New Classes of Compactifications of $d = 11$ Supergravity,}
Class.~ Quant.~Grav.~{\bf 2} (1985) L1.}
\lref\DiLithium{B.\ de Wit, H.\ Nicolai,  \nihil{A New $SO(7)$
Invariant
Solution of $d = 11$ Supergravity,}  Phys.~ Lett.~{\bf 148B} (1984) 60;
\hfill \break  P.~van~Nieuwenhuizen and N.P.~Warner, \nihil{New
Compactifications of Ten and Eleven Dimensional Supergravity on
Manifolds which are not Direct Products,}
Commun. Math. Phys.~{\bf 99} (1985) 141.}
\lref\FFZ{S.\ Ferrara, C.\ Fronsdal and A.\ Zaffaroni,
 \nihil{On N=8 supergravity on AdS(5) and N=4 superconformal
Yang--Mills
theory,} Nucl.~Phys.~{\bf B532} (1998) 153, \eprt{hep-th/9802203}.}
\lref\SupCurr{E.\ Bergshoeff, M.\ de Roo and B.\ de Wit,
\nihil{Extended
Conformal  Supergravity,} Nucl. Phys. {\bf B182} (1981) 173; \hfill
\break
P.\ Howe, K.S.\ Stelle, P.K.\ Townsend,  \nihil{Supercurrents,}
Nucl.~Phys.~{\bf B192} (1981) 332.}
\lref\LJR{L.J.\ Romans, \nihil{New Compactifications of Chiral $N=2$,
$d=10$ Supergravity,}  Phys.~Lett.~{\bf 153B} (1985) 392.}
\lref\MetAns{B.\ de Wit and H.\ Nicolai, \nihil{On the Relation
Between $d=4$ and $d=11$ Supergravity,} Nucl.~Phys.~{\bf B243}
(1984) 91; \hfil \break
B.\ de Wit, H.\ Nicolai and N.P.\ Warner,
\nihil{The Embedding of Gauged $N=8$ Supergravity into $d=11$
Supergravity,}
Nucl.~Phys.~{\bf B255} (1985) 29.}
\lref\GPPZ{L.\ Girardello, M.\ Petrini, M.\ Porrati and
A.\ Zaffaroni,  \nihil{Novel Local CFT and Exact Results on
Perturbations
of $N=4$ Super Yang--Mills from AdS Dynamics,}  \eprt{hep-th/9810126}.}
\lref\JDFZ{J.\ Distler and F.\ Zamora,
\nihil{Nonsupersymmetric Conformal Field Theories from Stable Anti-de
Sitter Spaces,}  \eprt{hep-th/9810206}.}
\lref\PBDF{P.\ Breitenlohner and D.Z.\ Freedman,
\nihil{Positive
Energy in Anti-de Sitter Backgrounds and Gauged Extended Supergravity,}
Phys.~ Lett.~{\bf 115B} (1982) 197; \nihil{Stability  in Gauged
Extended
Supergravity,}  Ann.~Phys.~{\bf 144} (1982) 249.}
\lref\PosEn{L.\ F.\ Abbott and S.\ Deser,  \nihil{Stabilty of Gravity
with
a Cosmological Constant,}  Nucl.~Phys.~{\bf B195} (1982) 76;
\hfil\break
G.W.\ Gibbons, C.M.\ Hull and N.\ P.\ Warner,  \nihil{The Stability
of Gauged Supergravity}  Nucl.~Phys.~{\bf B218} (1983) 173;
\hfil\break
W.\ Boucher, \nihil{Positive Energy without Supersymmetry,}
Nucl.~Phys.~{\bf B242} (1984) 282; \hfil \break
L.\ Mezincescu and P.K.\ Townsend, \nihil{Positive Energy and the
Scalar
Potential in Higher Dimensional (Super)Gravity Theories,}
Phys.~Lett.~{\bf 148B} (1984) 55; \hfil \break
L.\ Mezincescu and P.K.\ Townsend, \nihil{Stability  at a Local
Maximum
in Higher Dimensional Anti-de Sitter Space and Applications to
Supergravity,}
Ann.~Phys.~{\bf 160} (1985) 406.}
\lref\ConfDims{S.S.\ Gubser, I.R.\ Klebanov, A.M.\ Polyakov,
\nihil{Gauge Theory  Correlators from Non-Critical String Theory,}
Phys.~Lett.~{\bf B428}  (1998) 105, \eprt{hep-th/9802109}; \hfil
\break
E.\ Witten, \nihil{Anti-de Sitter space and holography,}
Adv. Theor.  Math.  Phys. {\bf 2} (1998) 253, \eprt{hep-th/9802150}.}
\lref\SSG{S.S.\ Gubser, \nihil{Einstein Manifolds and Conformal
Field Theories,}  \eprt{hep-th/9807164}}
\lref\MHKS{M.\ Henningson, K.\ Skenderis,  \nihil{The Holographic Weyl
Anomaly,}  J.~High Energy Phys.~{\bf 9807} (1998) 023,
\eprt{hep-th/9806087}.}
\lref\OPEfourD{D.\ Anselmi, M.\ Grisaru and A.\ Johansen,
\nihil{A Critical Behavior of Anomalous Currents, Electric--Magnetic
Universality and CFT in Four Dimensions,} Nucl.~Phys.~{\bf B491}
(1997) 221-248, \eprt{hep-th/9601023};  \hfil \break
D.\ Anselmi, D.\ Z.\ Freedman, M.\ T.\ Grisaru, A.\ A.\ Johansen,
\nihil{Universality of the Operator Product Expansions of SCFT
in Four Dimensions,} Phys.~Lett.~{\bf B394} (1997) 329-336,
\eprt{hep-th/9608125};  \hfil \break
D.\ Anselmi, \nihil{The N=4 Quantum Conformal Algebra,}
\eprt{hep-th/9809192}; \nihil{Quantum Conformal Algebras and Closed
Conformal
Field Theory,} \eprt{hep-th/9811149}.}
\lref\BdWHN{B.\ de Wit, H.\ Nicolai, \nihil{The Consistency of the
$S^7$  Truncation in $d = 11$ Supergravity,} Nucl.~Phys.~{\bf B281}
(1987) 211.}
\lref\IRKEW{I.R.\ Klebanov and E. Witten, \nihil{Superconformal
Field
Theory on Three-Branes at a Calabi-Yau Singularity,}
\eprt{hep-th/9807080}.}
\lref\Attractors{S.\ Ferrara, R.\ Kallosh and A.\ Strominger,
\nihil{$N=2$ Extremal Black Holes,} Phys.  Rev. {\bf D52} (1995)
5412-5416,
\eprt{hep-th/9508072};  \hfil \break
A.\ Strominger, \nihil{Macroscopic Entropy of $N=2$ Extremal Black
Holes,}
Phys.~ Lett. {\bf B383} (1996) 39-43, \eprt{hep-th/9602111};  \hfil
\break
S.\ Ferrara and R.\ Kallosh, \nihil{Supersymmetry and
Attractors,}
Phys.  Rev. {\bf D54} (1996) 1514-1524,  \eprt{hep-th/9602136};  \hfil
\break
G.\ Moore, \nihil{Arithmetic and Attractors,}
\eprt{hep-th/9807056};
\nihil{Attractors and Arithmetic,} \eprt{hep-th/9807087}.}
\lref\SKES{S.\ Kachru and E.\ Silverstein,
\nihil{4-D Conformal Theories and Strings on Orbifolds,}
Phys. Rev. Lett.{\bf 80} (1998) 4855, \eprt{hep-th/9802183}.}
\lref\NS{N.\ Seiberg, \nihil{Electric--Magnetic Duality in
Supersymmetric
Non--Abelian Gauge Theories,} Nucl.~Phys.~{\bf B435} (1995) 129,
\eprt{hep-th/9411149}.}
\lref\JMalda{J.~Maldacena, \nihil{The Large $N$ Limit of Superconformal
Field Theories and Supergravity,}, Adv.~Theor. Math. Phys.~{\bf 2}
(1998) 231 \eprt{hep-th/9711200}.}
\lref\LogPer{ Y.\ Huang, H.\ Saleur, C.G.\ Sammis
and D.\ Sornette,   \nihil{Precursors, Aftershocks, Criticality and
Self-Organized  Criticality,} cond-mat/9612065; \hfil\break
H.\ Saleur, C.G.\ Sammis and D.\ Sornette,
J. Geophys. Res 101 (1996) 17661; \hfil\break
Al.B.\ Zamolodchikov, \nihil{Resonance Factorized Scattering
and Roaming Trajectories,} ENS-LPS-335.}
%
%
\Title{\vbox{
\hbox{CERN-TH/98-387}
\hbox{USC-98/18}
\hbox{\tt hep-th/9812035}
}}
{\vbox{\vskip -1.0cm
\centerline{\hbox{New Vacua of Gauged $\cN=8$ Supergravity}}
\vskip 8 pt
\centerline{ \hbox{in Five Dimensions}}}}
\vskip -.3cm
\centerline{Alexei Khavaev, Krzysztof Pilch }
\medskip
\centerline{{\it Department of Physics and Astronomy}}
\centerline{{\it University of Southern California}}
\centerline{{\it Los Angeles, CA 90089-0484, USA}}
\medskip
\centerline{and }
\medskip
\centerline{Nicholas P.\ Warner}
\medskip
\centerline{{\it Theory Division, CERN \footnote{*}{\rm On
leave from Department of Physics and Astronomy,
USC,  Los Angeles, CA 90089-0484} }}
\centerline{{\it CH-1211 Geneva 23, Switzerland}}

\bigskip
\bigskip
We analyze a particular $SU(2)$ invariant sector of the scalar
manifold of gauged $\cN=8$ supergravity in five dimensions, and find
all the critical points of the potential within this sector.  The
critical points give rise to Anti-de Sitter vacua, and preserve at
least an $SU(2)$ gauge symmetry.  Consistent truncation implies that
these solutions correspond to Anti-de Sitter compactifications of IIB
supergravity, and hence to possible near-horizon geometries of
$3$-branes.  Thus we find new conformal phases of softly broken
$\cN=4$ Yang--Mills theory.  One of the critical points preserves
$\cN=2$ supersymmetry in the bulk and is therefore completely stable,
and corresponds to an $\cN=1$ superconformal fixed point of the
Yang--Mills theory.  The corresponding renormalization group flow from
the $\cN=4$ point has $c_{{\rm IR}}/c_{{\rm UV}} = 27/32$.  We also
discuss the ten-dimensional geometries corresponding to these critical
points.

\vskip .3in


\Date{\sl {December, 1998}}

%
\parskip=4pt plus 15pt minus 1pt
\baselineskip=15pt plus 2pt minus 1pt
%
\newsec{Introduction}

The correspondence between AdS supergravity theories and
superconformal field theories on branes has been examined from many
perspectives, and this has led to a much deeper and richer
understanding of how these correspondences work.  It is our purpose in
this letter to re-examine some of the issues that were important in
gauged supergravity 15 years ago, but now considered from the
perspective of superconformal Yang-Mills theories.  We consider
perhaps the best substantiated correspondence \JMalda: that of gauged
$\cN=8$ supergravity in five dimensions \refs{\GRW,\PPvN} and $\cN=4$
supersymmetric Yang-Mills theory on $3$-branes.  In particular we
analyse the potential in this supergravity model, finding a class of
critical points that have at least $SU(2)$ gauge symmetry in the
supergravity (or $R$-symmetry of the Yang-Mills theory).  This class
includes a non-trivial supersymmetric critical point.

Before proceeding with the analysis we review some of the relevant
ancient history.  In the early 1980's much work was done on testing
and establishing that the maximal gauged supergravities were indeed
{\it embedded} in the sphere compactifications of  various higher
dimensional theories.  By ``embedded'' we mean that the full
non-linear gauged supergravity action can be fully encoded in the
action
and field equations of the higher dimensional theory, and in
particular, a solution of the gauged supergravity theory can be
precisely mapped onto a solution of the higher dimensional theory.
The possibility of such a consistent truncation was considered quite
remarkable in that the states of the lower dimensional gauged
supergravity involved non-trivial spherical harmonics in the higher
dimensional theory, and it seemed that many miraculous identities
would be needed if the full non-linearities of the gauged supergravity
were going to decouple consistently from the higher Kaluza-Klein
states.  This was most extensively studied for the $S^7$
compactification of eleven dimensional supergravity to $\cN=8$ gauged
supergravity in four dimensions, and a vast body of evidence was
assembled in support of ``consistent truncation,'' and ultimately the
full non-linear embedding was explicitly constructed \BdWHN.  Less is
known about the embedding of other maximal theories in other
dimensions, and the complete Ans\"atze were never constructed.
However, given the proof in \BdWHN, and the structural similarities of
the various maximal gauged theories, particularly between
the five-dimensional and the four-dimensional $\cN=8$ theories, it
seems extremely likely that such theories are embedded in their higher
dimensional counterparts.  This is further supported by quite a number
of non-trivial consistency checks that have been performed over the
years through the construction of explicit solutions.  We will thus
take it as a given that the five-dimensional $\cN=8$ theory is
embedded in the $S^5$ compactification of IIB supergravity.

The consistency of the truncation has a simple, and fundamental
meaning for the Yang-Mills theory on the $3$-branes.  As is now fairly
well established, the supergravity scalars represent couplings of the
relevant and marginal chiral primary perturbations of the $\cN=4$
Yang-Mills theory.  These operators constitute a very particular
subset of all of the relevant and marginal perturbations of the
theory, namely those that belong to the ``short'' $\cN =4$ multiplet
of the energy momentum tensor \refs{\FFZ,\SupCurr}.  Consistent
truncation means that, at least for large $N$, this subset of chiral,
primary operators should close under operator product, as has been
discussed in \OPEfourD.  Such a closed operator algebra also means
that if one turns on couplings to these chiral primary operators, then
the renormalization group flow should be determined entirely by these
relevant (and marginal) operators chiral primaries, and not by the
``irrelevant'' higher Kaluza-Klein states.  One can implement this
very explicitly within the supergravity theory: If one can find a
critical point of the scalar potential, $\cP$, then this corresponds
to a solution of the ten-dimensional theory, and hence to a ``phase''
of the $3$-branes. There is of course the $\cN=8$ supersymmetric,
$SO(6)$ invariant critical point, corresponding to the $S^5$
compactification of the $IIB$ supergravity, but there are also other
critical points.  These ``new'' critical points are generically at
negative values of $\cP$, and so there are solutions in Anti-de Sitter
space, and the corresponding $3$-brane field theories are thus
conformal.  One can make this correspondence even more explicit
\refs{\GPPZ,\JDFZ} by constructing ``interpolating'' solutions: that
is solutions that, at large distance from the branes, are at the
maximally symmetric critical point and then flow, as the distance from
the brane decreases, to another critical point.  Since the distance
from the brane represents the scale in the Yang-Mills theory, such
interpolating solutions must represent explicit renormalization group
flows from the U.V.  fixed point ($\cN=4$ Yang-Mills theory) to a new
conformal I.R. phase of the Yang-Mills theory.  Consistent truncation
means that this flow will be entirely determined by the equations of
motion of supergravity in five dimensions.

In the gauged supergravity theories there was also the concern that
since one was dealing with critical points of a potential, stability
of solutions is an issue.  However, because of the Planck scale of the
potential there is a strong gravitational back-reaction that tends to
stabilize some of the naively unstable critical points.  There is the
Breitenlohner-Freedman condition \refs{\PBDF,\PosEn}, that states that
a scalar wave equation $\square \phi - \alpha \phi$ is perturbatively
stable in an AdS space of dimension $d$ and radius $R$ if $\alpha < {1
\over 4 R^2} (d-1)^2$.  There is also a more powerful, but less
general result that states that if a solution has any supersymmetry
then the solution is completely semi-classically stable \PosEn.  The
statement of stability from the point of view of the phases of
Yang-Mills appears to be a statement of unitarity.  This is most
directly seen from the connection between the conformal dimension of
an operator in the Yang-Mills theory and the ``mass'' of a
supergravity scalar \ConfDims.  Scalars that do not satisfy the
Breitenlohner-Freedman bound correspond to operators with complex
conformal dimensions \refs{\GPPZ,\JDFZ} and thus represent non-unitary
perturbations.  Such perturbations have been considered in field
theories in $1+1$-dimensions, and in systems with self-organized
criticality, where they lead to ``log-periodic'' behaviour, and
``roaming'' renormalization group flows (see, for example,
\LogPer). In the Yang-Mills theory on the $3$-brane, all the
``unstable'' operators have a dimension whose real part is $2$, and
hence they are relevant and drive an ``oscillatory flow.'' The
physical interpretation of such operators in field theory however is
far from clear.

Another possible issue of concern is that a critical point
of the supergravity theory might be a pathology of large
$N$ gauge theories.  While such large $N$ pathologies
may still be interesting in their own right, it would
be nice to know that a given critical point represents
a solution at finite $N$.  One way to ensure this is
if a solution represents a solution to the string theory,
and this is much more likely if the solution is stable,
and indeed if it is supersymmetric.

Thus, a classification of critical points of the potential
of gauged $\cN=8$ supergravity in five dimensions takes on
new relevance since it represents a classification of
infra-red fixed points of large N, $\cN=4$ Yang Mills theory.
Moreover, the supersymmetric
critical points are particularly significant since they
represent very stable fixed points, that should be
present even in $\cN =4$, $SU(N)$ Yang-Mills with finite $N$.

The scalar potential of gauged $\cN=8$ supergravity in five
dimensions is a function of $42$ scalars.  It is invariant under the
$R$-symmetry, $SO(6)$, and under the $SL(2,R)$ symmetry of
the original IIB  theory in ten dimensions.   This means that
the potential is a function of $24$ independent variables, and
in terms of physical scalars the potential has two flat directions
at every point, coming from the non-compact generators of $SL(2,R)$.
This number of parameters is too large to be practicably managed,
and so we reduce the problem by seeking out all critical points
that reduce the gauge/$R$-symmetry to a group containing
a particular $SU(2)$ subgroup of $SO(6)$.
This choice is not only motivated by the pragmatic consideration
of actually being able to perform the calculation, but also by
the fact that $SU(2)$ is the $R$-symmetry of $\cN=2$ gauge theories.
We do not actually find a solution with $\cN=2$ supersymmetry on
the brane, but this reduction
of the problem to the $SU(2)$ invariant subsector is still wide
enough to enable us to find one critical point that has
$\cN=1$  supersymmetry on the brane, {\it i.e.} $\cN=2$ supersymmetry
in the supergravity theory.

Finally there is the value of the
cosmological constant at a critical point.  It is a consequence
of the work of \refs{\MHKS,\SSG} that this translates directly into
the central charge of the conformal theory.  We compute the
central charges at the various critical points, and find that
the supersymmetric critical point has $c_{{\rm IR}}/c_{{\rm UV}} =
27/32$.

In section 2 we discuss the truncation of the scalar manifold
and potential to a particular $SU(2)$ invariant subsector.
In section 3 we describe, and catalogue properties of, all the
critical points in this invariant subsector.  In section 4 we
discuss some of the implications of our results.  We focus on the
non-trivial supersymmetric critical point, and describe the geometry
of its embedding into the ten-dimensional theory. Indeed, we use the
close  similarity between the five dimensional and the four
dimensional $\cN=8$ theories to infer the proper generalization
of \MetAns, and conjecture the full non-linear Ansatz for the
``internal'' compactifying metric for gauged $\cN=8$ supergravity
in five dimensions.

\newsec{Constructing the potential on the $SU(2)$ invariant sector}

As was described in \NWcpts, an effective way of searching for
interesting subsets of critical points of the potential is to restrict
the problem to the space, $\cS$, of singlets of some invariance group,
$G$.  It is a trivial consequence of Schur's lemma that any variation
of the potential, $\cP$, about $\cS$ in a non-singlet direction is
necessarily quadratic, and thus any critical point of $\cP$ on $\cS$
is necessarily a critical point on the whole space of scalars.  Here
we consider $SU(2)_\cR$ subgroups\foot{The subscript $\cR$ is to
distinguish this $SU(2)$ from various other $SU(2)$ subgroups of
$E_{6(6)}$.} of the $\cR$-symmetry $SU(4)$.  There are four distinct
such subgroups in $SU(4)$ and these can be characterized in terms of
how the ${\bf 4}$ of $SU(4)$ decomposes under $SU(2)$.  Specifically,
one can have: (i) ${\bf 4} \to {\bf 2} \oplus {\bf 1} \oplus {\bf 1}$,
(ii) ${\bf 4} \to {\bf 2} \oplus {\bf 2}$, (iii) ${\bf 4} \to {\bf 3}
\oplus {\bf 1}$, or (iv) ${\bf 4} \to {\bf 4}$.  Only the first
possibility will be analysed in this letter since it has the largest
singlet structure in $E_{6(6)}$ and is thus most likely to yield new
critical points.

Our choice of $SU(2)_\cR$ commutes with an $H_0 \equiv SU(2)
\times U(1)$ in $SU(4)$.  The commutant of $SU(2)_\cR$ is extended to
$H_1 \equiv SU(2) \times SL(2,\IR) \times \IR^+$ in
$SL(6,\IR)$,  and finally  to $H_2\equiv SO(5,2) \times \IR^+$ in
$E_{6(6)}$.
\foot{In writing $\IR^+$ here we are dropping two discrete $\ZZ_2$
subgroups: One generated by $-1 \in SL(6,\IR)$, and the other
by ${\rm diag}(-1,-1,-1,-1,1,1)$. These $\ZZ_2$'s are  symmetries
of the potential and so all our critical points will, in fact,
come in $\ZZ_2 \times \ZZ_2$ multiplets.}   If one thinks
of the obvious $SO(3) \times SO(2,2)$ subgroup of $SO(5,2) \times
\IR^+$, and recalls that $SO(2,2) = SL(2,\IR) \times SL(2,\IR)$ then
one of these factors is the $SL(2,\IR)$ of $H_1$, while the other
$SL(2,\IR)$ factor is the scalar manifold of the ten-dimensional
theory. The $SO(3)$ factor above is the $SU(2)$ of $H_0$. As one would
expect, the maximal compact subgroup $SO(5) \times SO(2)$ of $SO(5,2)$
is the subgroup of $USp(8)$ that commutes with $SU(2)_\cR$. Thus the
manifold of scalar singlets is given by
\eqn\singman{\cS ~=~ {SO(5,2) \over SO(5) \times SO(2)} \times \IR^+
\ .}

To find a simple parametrization of the potential on this space
we need to fix the invariances as cleanly as possible.  The
manifold, $\cS$, is $11$-dimensional, but the potential
has a residual $H_{inv} \equiv H_0  \times SL(2,\IR)
\equiv SU(2) \times U(1) \times SL(2,\IR)$
invariance,  which has seven parameters.  This means that we should be
able to reduce the potential to a function of four variables,
and indeed we can.  We represent an element of the coset, $\cS$, by
$\rho=e^\alpha \in \IR^+$ and by the exponential of a
$7 \times 7$ matrix,  $M$, with  $M_{i6}= M_{6i} = x_i$;
$M_{i7}= M_{7i} = y_i$, $i =1,\dots,5$, and all other
$M_{ij}$ set to zero.  We now argue that there is a gauge in
which we can take $x_1=x_3=x_5=0$ and $y_i=0, i=1,3,4,5$ ,
leaving four parameters: $x_1,y_2,x_4$ and $\alpha$.

{}First the $SU(2)$ of $H_{inv}$ acts as the triplet of $x_i$ and
$y_i$,  $i=1,2,3$. We could completely fix this $SO(3)$ by setting
$x_3=x_2=0$ and $y_3=0$, but we start by partially fixing it by
setting  only  $x_3=y_3=0$.  This leaves an $SO(2)$ subgroup which
may be  thought of as acting on the matrix $M_0=\left(
\matrix{x_1&y_1\cr  x_2&y_2\cr}\right)$ from the left.
A linear combination of the $U(1)$
and the $SO(2)$ subgroup of $SL(2,\IR)$ acts on $M_0$ by right
multiplication.  These two independent $SO(2)$ actions
can be used to diagonalize $M_0$.  We have thus set
$x_2 = y_1=x_3=y_3=0$, and are left with the other linear
combination of the $U(1)$ and the $SO(2)$ subgroup of $SL(2,\IR)$, and
the non-compact generators of $SL(2,\IR)$.  We can fix the latter
by setting $y_4=y_5=0$, while the remaining $SO(2)$ rotates the
$(4,5)$ coordinates and so can be used to set $x_5 = 0$, which
completes the gauge fixing.

To constuct the potential we we need to construct the
$27 \times 27$ $E_{6(6)}$ matrix, $\cV$.  Under
$SU(2)_\cR \times SO(5,2)\times R^+$ the ${\bf 27}$ of $E_{6(6)}$
decomposes as:
$$
{\bf 27} ~\to~ {\bf (1,1)}(+4)
{}~\oplus~ {\bf(1,7)}(-2)~\oplus ~{\bf (2,8)}(+1)~ \oplus~
{\bf (3,1)}(-2) \ .
$$
This means that $\cV$, when written in the proper basis
consists of: (i) the  exponential $M$ and multiplied
by $\rho^{-2}$, (ii) two copies of the  exponential of
the spinor representative of $M$,  multiplied them by $\rho$, and (iii)
the matrix ${\rm diag}(\rho^4,\rho^{-2},\rho^{-2},\rho^{-2})$.
To assemble the potential one then needs to rotate this back into
the $SL(6,\IR) \times SL(2,\IR)$ basis.
The exponentials of the gauge fixed matrices are elementary to compute,
and the basis rotations are tedious, but straightforward.
Using ${\it Mathematica}^{TM}$ we assembled all of this into the
potential as described in \GRW\ and obtained:
\eqn\potredcd{\eqalign{
\cP	~=~ & {g^2 \over 32}~\rho^{-4}~ \big( \cosh(4 r_x) \cosh(4 r_y) -
\cosh(4 r_x) - \cosh(4 r_y) \cr & + 4 \cos(2 \theta)  \sinh(2 r_x)^2
\sinh(2 r_y)^2 - 7\big) ~-~ {g^2 \over 2}~ \rho^2 \cosh(2 r_x)
\cosh(2 r_y) \cr & +~ {g^2 \over 64} ~\rho^8 ~\big(\cosh(4 r_x) +
2 \cosh(4 r_x ) \cosh(4 r_y) - 2 \cos(2 \theta)
\sinh(2 r_x)^2 - 3 \big)
\ ,}}
where $x_1 =  r_x \cos\theta$, $x_4 =  r_x\sin\theta$, $\rho =
e^\alpha$ and $r_y = y_2$.

\newsec{Critical Points}

It is elementary to find the critical points of \potredcd, and
some of the details are summarized in Table 1.
As discussed in \GRW, one can determine the number of unbroken
supersymmetries at a given critical point by finding the eigenvalues,
$\mu_i$, of the tensor $W_{ab}$.  The number of
supersymmetries is equal to the number of $\mu_i$ for which $|\mu_i| =
\sqrt{- 3 \Lambda/g^2}$, where $\Lambda = \cP$ is the cosmological
constant at the critical point.  The critical points and
their supersymmetries are as follows:

\item{(i)} First there is the well known, trivial critical point at
which  all the scalars vanish, and whose cosmological constant is
$\Lambda = - 3 g^2/4$, and which preserves $\cN=8$ supersymmetry.

\item{(ii)}  There is a critical point at $ x_1 = y_2  = 0$,
$x_4 = {1 \over 4} \log(3)$, and $\alpha = {1 \over 12} \log(3)$,
and the cosmological  constant is $\Lambda = - {3^{5/3} \over 8}g^2$.
This scalar vev actually lives in the $SL(6,\IR)$ subgroup of
$E_{6(6)}$, and it is $SO(5)$ invariant. This critical point
was found in \GRW. The eigenvalues of  $W_{ab}$ are
all equal to $-2. 3^{-1/6}$, and hence there is no
supersymmetry at this point.  It was also shown in \JDFZ\ that
this critical point is perturbatively unstable.

\item{(iii)}  There is a critical point at $ x_1 = y_2  =
 \pm {1 \over 4} \log(2 - \sqrt{3})$, $x_4 = 0$, and
$\alpha = 0$, and the cosmological
constant is $\Lambda = - {27 \over 32}g^2 $.
This scalar vev corresponds to  the $SU(3)$ invariant critical point
discovered in \GRW.   The eigenvalues of  $W_{ab}$ are
$-{7 \over 4}$ and  $-{9 \over 4}$ with multiplicities
of $6$ and $2$ respectively, and so there is no
supersymmetry.

\item{(iv)}  There is a critical point at $ x_1 = x_4  = 0$,
$y_2 = \pm {1 \over 4}
\log({1\over 5}(11 - 4 \sqrt{6}))$, and $\alpha = {1 \over 12}
\log(10)$, and the cosmological
constant is $\Lambda = - {3 \over 8}({25 \over 2})^{1/3} g^2 $.
The non-zero vev of $\alpha$  reduces the $SO(6)$ invariance
to $SU(2) \times SU(2) \times U(1)$, and the  non-zero vev of $y_2$
further reduces this to  $SU(2) \times U(1) \times U(1)$.
The eigenvalues of  $W_{ab}$ are $-3.10^{-1/6}$ and  $-9.10^{-2/3}$
each with a multiplicity of $4$, and once again there is no
supersymmetry.

\item{(v)}  There is a critical point at $ x_1 = y_2 = \pm {1 \over 4}
\log(3)$, $x_4 = 0$, and $\alpha = {1 \over 6} \log(2)$,
and the cosmological constant is $\Lambda = - {2^{4/3} \over 3}g^2$. As
in (iii), the non-zero vev of $ x_1 = y_2$ reduces the $SO(6)$
invariance to $SU(3)$, and then the non-zero vev of $\alpha$ further
reduces this to $SU(2) \times U(1)$. The eigenvalues of $W_{ab}$ are
$-{7 \over 3}~2^{-1/3}$, $-{4 \over 3}~2^{2/3}$ and $- 2^{2/3}$ with a
multiplicities of $4,2$ and $2$ respectively. The last eigenvalue is
equal to $-\sqrt{- 3 \Lambda/g^2}$, and so this critical point has an
unbroken $\cN=2$ supersymmetry.

As discussed in the introduction, these critical points may be
thought of as infra-red fixed points of the Yang-Mills theory
on the branes.  Since all the cosmological constants are
negative, and therefore admit anti-de Sitter metrics, the
corresponding gauge theories are conformal.  Using the results
of \refs{\MHKS,\SSG} one can compute the central charge of these
conformal theories.  Indeed, from \MHKS\ one sees
that the ratio of central charge, $c_{{\rm IR}}$, of the new fixed
point compared to $c_{{\rm UV}}$, the central charge of the $\cN=4$
symmetric fixed point is given by:
\eqn\cIRcUV{{c_{{\rm IR}} \over c_{{\rm UV}}} ~=~
\bigg({\Lambda_{{\rm IR}} \over
 \Lambda_{{\rm UV}}}\bigg)^{-{3\over 2}} ~=~  \bigg(- { 4 \Lambda_{{\rm
 IR}} \over
 3 g^2 }\bigg)^{-{3\over 2}}  \ ,}
where $\Lambda_{{\rm IR}}$ and $\Lambda_{{\rm UV}} = - {3 \over 4}
g^2$ are the cosmological constants at the corresponding critical
points.  This ratio of central charges is also given in Table 1.

\goodbreak

{\vbox{\ninepoint{
$$
\vbox{\offinterlineskip\tabskip=0pt
\halign{\strut\vrule#
&~$#$~\hfil\vrule
&~$#$~\hfil\vrule
&~$#$~\hfil\vrule
&~$#$~\hfil\vrule
&~$#$\hfil
&\vrule#
\cr
\noalign{\hrule}
&
{\rm Critical}
&
{\rm Unbroken\ Gauge}
&
{\rm Cosmological}
&
{\rm Unbroken}
&
{\rm Central Charge}
&\cr
&
{\rm Point}
&
{\rm Symmetry}
&
{\rm Constant}
&
{\rm Supersymmetry}
&
{\rm c}_{IR}/{\rm c}_{UV}
&\cr
\noalign{\hrule}
&
(i)
&
SO(6)
&
-{3 \over 4} g^2
&
\cN = 8
&
1
&\cr
&
(ii)
&
SO(5)
&
- {3^{5/3} \over 8} g^2
&
\cN=0
&
{2 \sqrt{2} \over 3} \sim 0.9428
&\cr
&
(iii)
&
SU(3)
&
-{27 \over 32} g^2
&
\cN =0
&
{16 \sqrt{2} \over 27} \sim 0.8381
&\cr
&
(iv)
&
SU(2) \times U(1) \times U(1)
&
- {3 \over 8}({25 \over 2})^{1/3} g^2
&
\cN=0
&
{4 \over 5}  = 0.8
&\cr
&
(v)
&
SU(2) \times U(1)
&
-{2^{4/3} \over 3}g^2
&
\cN =2
&
{27 \over 32} \sim 0.8438
&\cr
\noalign{\hrule}}
\hrule}$$
\vskip-10pt
\noindent{\bf Table 1:}
{\sl
The critical points of the potential \potredcd.  The
unbroken gauge symmetry is that of the supergravity, and
corresponds to the $\cR$-symmetry on the branes.
The unbroken supersymmetry is that of the supergravity theory,
and should be halved to get the supersymmetry on the branes.}
\vskip10pt}}}

\newsec{Discussion}

The critical points described above give rise to new phases of
Yang--Mills theory that are presumably strongly coupled, and so it
is not altogether obvious what their massless spectra should be.
Indeed, the previously known, non-trivial  critical points ((ii) and
(iii) in Table 1) were discussed in \refs{\GPPZ,\JDFZ}, and have an
irrational central charge ratio, and so cannot have a spectrum that is
simply related to the perturbative spectrum of the $\cN=4$ Yang-Mills
theory. The two new critical points found in this letter
((iv) and (v) in Table 1) may ultimately prove a little more amenable.
Both have rational central charge, and one of them has $\cN=1$
supersymmetry and is therefore very stable.

Before discussing the supersymmetric critical point in some
detail, we wish to note some possible patterns in our albeit very
small amount of data.  First, we note that the central charge
decreases with the amount of global symmetry (the central charge
jumps back up if there is residual supersymmetry).  Secondly, both
in five dimensions and in four dimensions \NWcpts, we note that there
is either no critical point, or a unique critical point with a
given global symmetry.  This is similar in spirit  to the
results of \Attractors\ which suggest that the near horizon geometry
of a brane configuration should be uniquely determined
by the angular momentum, or horizon symmetry, and charges of
the branes.

There is an apparent parallel between our supersymmetric
critical point, and the $\cN=1$ supersymmetric model considered in
\refs{\IRKEW,\SSG}.   The latter model was obtained by first doing a
$\ZZ_2$ orbifold of the $\cN=4$ model so as to reduce it to $\cN =2$,
and then turning on a relevant perturbation in the field theory to
break it to an $\cN=1$ model and flow to a new fixed point. (This
fixed point has a marginal perturbation, and the resulting fixed line
connects to the conformal window of \NS.)  The $\ZZ_2$ orbifold is
made as in \SKES: The $\ZZ_2$  acts on $S_5$ via the matrix $\sigma
={\rm diag}(-1,-1,-1,-1,1,1)$ multiplying the Euclidean coordinates of
$\IR^6$.  There is also a simultaneous action on the branes in which
the branes are separated into two groups of $N$, and then these groups
are interchanged. The relevant perturbation that breaks to $\cN=1$ in
\refs{\IRKEW,\SSG} is in the twisted sector of the theory and involves
turning on a fermion masses so as to reduce the $\cR$-symmetry from
$SU(2) \times U(1)$ to $U(1)$.

The $\cN=1$ supersymmetric critical point found here involves a
two parameter submanifold, $\cS_0$, of two {\it commuting} scalars in
$E_{6(6)}/USp(8)$. These are parametrized by $\alpha$ and $\beta =
x_1=y_2$, and are characterized respectively as (i) matrices in
$SL(6,\IR)$ of the form $M(\alpha) = {\rm diag} (e^\alpha, e^\alpha,
e^\alpha, e^\alpha, e^{-2\alpha},e^{-2\alpha})$, and (ii) the
non-compact generator of $E_{6(6)}$ that breaks $SO(6)$ down to $SU(3)$
(an explicit expression for this was given in \GRW). First note that
the orbifold generator, $\sigma$, is the same as $M(\alpha = i \pi)$,
and also that $\sigma$ commutes with the scalars on $\cS_0$. The
consequence of the first observation is that $\sigma$ and $M$ make the
same fields massive. The consequence of the second observation is that
the scalar manifold $\cS_0$ is part of the scalar manifold of the
untwisted sector of the orbifold theory of \refs{\IRKEW,\SSG}.
Turning on $\alpha$
breaks $SO(6)$ to $SU(2) \times SU(2) \times U(1)$, and turning on
$\beta$ reduces this to $SU(2) \times U(1)$, where the $U(1)$ becomes
the $\cR$-symmetry at the $\cN=1$ supersymmetric point, and $SU(2)$ is
an additional global symmetry. From the AdS/CFT correspondence, turning
on $\beta$ corresponds to turning on very specific fermion masses
\ConfDims. While this relevant operator is not the same as that
used in \IRKEW, it is possible that the IR fixed points of the
renormalization group flows could be related.

One can make the parallel  much closer by going to the
$\sigma$-invariant sector of the gauged $\cN=8$ supergravity
theory.  This is a truncation to a gauged $\cN=4$ supergravity
theory, and corresponds to a subsector of the untwisted sector
of the $\ZZ_2$ orbifold. (This $\sigma$-invariant sector does
{\it not} contain the hypermultiplets arising from the interchange
of the branes.)  The scalar submanifold $\cS_0$, and the
corresponding restriction of scalar potential survives this
truncation, and so the corresponding relevant perturbations of
the full orbifold theory will be described by this part of
the supergravity action.  Turning on the scalar vevs considered
in this paper will generate a superpotential for the chiral
multiplets, but a  different one from that described in \IRKEW\
(and it will have a smaller global symmetry).  The flow to the
critical point will then lead to a different fixed line.
Whether this fixed line
and that of \IRKEW\ are connected remains to be seen, but an
intriguing, though indirect piece  of evidence for such a
connection is the fact that both
renormalization group flows have $c_{{\rm IR}}/c_{{\rm UV}} = 27/32$.

Finally, it is instructive to consider the geometry of the
compactification that corresponds to our superconformal
critical point.

The full embedding of five dimensional, gauged $\cN=8$ supergravity
into the ten-dimensional theory has never been explictly written
down, but it is rather easy to infer part of it from the
the corresponding results for four-dimensional supergravity.
Turning on scalars in
$SL(6,\IR)$ corresponds to a modifying the metric of $S^5$
to that of a surface in $\IR^6$ defined by \MetAns\
\eqn\squish{\vec x^T \cdot S^T S\cdot \vec x ~=~ r^2 \ ,}
where $\vec x$ are cartesian coordinates on $\IR^6$, $S \in
SL(6,\IR)$, and $r$ is a constant.  This deformation is accompanied
by the introduction of ``warp-factors'' \DiLithium\ that are fractional
powers of $\mu = (\vec x^T \cdot( S^T S)^2 \cdot \vec x)$
in front of the metric of \squish\ and the metric the anti-de Sitter
space time. This means that turning on
$\alpha$ deforms the internal metric to a ellipsoid in a manner
reminiscent of rotating branes.

To lowest order the mode expansion of $S^5$ the perturbation
by $\beta$ corresponds to turning on
an $SU(3)$-invariant configuration for the anti-symmetric tensor
field $G_{mnp}$ in ten dimensions.  From what is known about
the $SU(3)$ invariant critical point \GRW\ and the corresponding
compactification  \LJR, we know that one should consider the
$S^5$ (and the deformation described above) as an $S^1$ bundle over
a (deformed) $\IC \IP^2$.  Let $\chi$ denote the fiber coordinate, and
let $K_{ij}$ be the holomorphic $2$-form on the base -- the latter is
not globally well defined, but the following Ansatz for $G_{mnp}$ is
\refs{\CNPNW,\LJR}:
\eqn\tensAns{G_{ij\chi} ~=~ a~e^{2 i \chi}~ K_{ij}\ , \qquad
G^*_{ij\chi} ~=~ a^*~e^{-2 i \chi}~ K_{\bar i \bar j} \ ,}
where $a$ is a constant and  $K_{\bar i \bar j}$ is the complex
conjugate of $K_{ij}$.  At higher orders in the perturbation
$\beta$, the metric of the $S^1$ fibration is further deformed
by a stretching of the $S^1$ fiber compared to the scale of the
base.  Putting this all together, the background will
be an $S^1$ fibration over a complex $2$-fold which itself
consists of a squashed $\IC \IP^2$.  The latter squashing
can be accomplished  by deforming the scale of the complex
coordinates so as to preserve an isometry,
$SU(2) \times U(1)$ of $\IC \IP^2$.  Indeed, in a properly
chosen coordinate patch one can arrange that this $SU(2) \times U(1)$
is the local isotropy group of the origin, with the remaining
translational isometries of $\IC \IP^2$ being broken.
A background tensor  field of the form \tensAns\ is then turned on.

For completeness sake we note that based on the arguments of
\MetAns,  a reasonable conjecture
for the full (inverse) metric on the internal space is given by:
\eqn\metansatz{\Delta^{-{2 \over 3}}~g^{mp} ~=~c~K^{m IJ}~K^{p KL}~
\widetilde \cV_{IJab}~\widetilde\cV_{KLcd}~\Omega^{ac}~\Omega^{bd} \ .}
In this equation $K^{m IJ} = -K^{m JI}$, $I,J =1,\dots,6$ are
the Killing fields on the $S^5$,  $\widetilde \cV_{IJab}$ is a
submatrix, defined in \GRW, of the inverse of the $27 \times 27$
scalar $E_{6(6)}$ matrix, $\cV$; $\Omega^{ab}$
is the $USp(8)$ symplectic invariant,  $\Delta =\sqrt{det(g_{mp})}$,
and $c$ is a normalization constant.  In addition to this,
the anti-de Sitter metric of the five-dimensional space time
must be rescaled by the warp-factor $\Delta^{-2/3}$.

The compactification that we have just outlined is
very different from the compactification considered in
\IRKEW.  Indeed, for the $\cN=1$ theory of
\IRKEW\  the relevant geometry at the conifold point was argued
to be that of the coset ${\cal T}^{1,1}$.  There is no background
tensor field.  Moreover, the critical
line described in \IRKEW\ involved the K\"ahler modulus of a blow up
of singularity.  If, after passing to a $\ZZ_2$ orbifold, our critical
point has a similar marginal deformation, then it  would be interesting
to see precisely to what it corresponds.  It seems very likely that it
would involve turning on the tensor field, which in terms of the
conifold, is suggestive of some kind of ``dual'' branch to the
construction of \IRKEW.

\goodbreak
\vskip2.cm\centerline{\bf Acknowledgements}
\noindent

N.W. would like to thank M.~Porrati and Y.~Oz for valuable discussions.
This work was supported in part
by funds provided by the DOE under grant number DE-FG03-84ER-40168.


\vskip 2cm

\listrefs

\vfill
\eject
\end